\documentclass[twocolumn,prb,showpacs]{revtex4}
\usepackage{bm,epsfig,subfigure,times,amsmath,dcolumn,amssymb}
\usepackage{graphicx,epstopdf}
\newcommand{\foo}{Li$_\text{4}$BN$_\text{3}$H$_\text{10}$}
\newcommand{\htwo}{H$_\text{2}$}
\newcommand{\ntwo}{N$_\text{2}$}
\newcommand{\nhthree}{NH$_\text{3}$}
\newcommand{\libn}{Li$_\text{3}$BN$_\text{2}$}
\newcommand{\lin}{Li$_\text{3}$N}
\newcommand{\amide}{LiNH$_\text{2}$}
\newcommand{\imide}{Li$_\text{2}$NH}
\newcommand{\libh}{LiBH$_\text{4}$}

\newcommand{\hvib}{$H_\text{vib}$}
\newcommand{\evib}{$E_\text{vib}$}
\newcommand{\svib}{$S_\text{vib}$} 
 
\begin{document}

\title{First principles study of the crystal structure and
  dehydrogenation pathways of Li$_\text{4}$BN$_\text{3}$H$_\text{10}$}

\author{Donald J. Siegel and C. Wolverton} 
\affiliation{Physical and Environmental Sciences Department, Ford Motor Company, MD3083/RIC, Dearborn, MI 48121}
\author{V. Ozoli\c{n}\v{s}}
\affiliation{Department of Materials Science and Engineering, University of California, Los Angeles, CA 90095}

\date{\today}

\begin{abstract}
  Using density functional theory we examine the crystal structure and
  the finite-temperature thermodynamics of formation and
  dehydrogenation for the new quaternary hydride \foo.  Two recent
  studies based on X-ray and neutron diffraction have reported three
  bcc crystal structures for this phase. While these structures
  possess identical space groups and similar lattice constants,
  internal coordinate differences result in bond length discrepancies
  as large as 0.2~\AA.  Geometry optimization calculations on the
  experimental structures reveal that the apparent discrepancies are
  an artifact of X-ray interactions with strong bond polarization; the
  relaxed structures are essentially identical.  Regarding reaction
  energetics, the present calculations predict that the formation
  reaction 3~LiNH$_\text{2}$ + LiBH$_\text{4}$ $\rightarrow$ \foo\ is
  exothermic with enthalpy $\Delta H^\text{T=300K} = -11.8$ kJ/(mol
  f.u.), consistent with reports of spontaneous \foo\ formation in the
  literature.  Calorimetry experiments have been reported for the
  dehydrogenation reaction, but have proven difficult to interpret.
  To help clarify the thermodynamics we evaluate the free energies of
  seventeen candidate dehydrogenation pathways over the temperature
  range $T = $ 0--1000~K.  At temperatures where H$_2$-release has
  been experimentally observed ($T \approx$ 520--630~K), the favored
  dehydrogenation reaction is \foo\ $\rightarrow$ \libn\ + \amide\ +
  4\htwo, which is weakly endothermic [$\Delta H^\text{T=550K} = 12.8$
  kJ/(mol \htwo)].  The small calculated $\Delta H$ is consistent with
  the unsuccessful attempts at re-hydriding reported in the
  literature, and suggests that the moderately high temperatures
  needed for H-desorption result from slow kinetics.
\end{abstract}

\maketitle

\section{Introduction}
Recent efforts to improve the efficiency and reduce the environmental
impact of automobile transport have focused on hydrogen-based fuel
cells (FC) and internal combustion engines (H$_2$ICE) as possible
replacements for current technologies powered by
fossil-fuels.\cite{nrc_hydrogen04} A significant obstacle to realizing
this transition is the on-board storage of hydrogen at high
gravimetric and volumetric densities.\cite{schlapbach01} To achieve
the storage densities necessary for mobile applications, novel means
for H-storage are necessary, and a set of targets have been
established to guide the search for new storage
systems.\cite{doe_targets} At present no known storage material or
mechanism meets these targets.

One promising avenue for efficient storage of hydrogen is via solid
state storage, such as in the form of complex- or metal-hydrides.
Solid storage has the advantage of providing volumetric densities
beyond what can be achieved with compressed \htwo\ gas or cryogenic
liquid storage. However, all known hydrides suffer from one or more of
the following limitations: low gravimetric densities, high
H-desorption temperatures, or an inability to be easily re-hydrided.
The search for new hydrides that overcome these limitations has
attracted intense interest during the past 5 years.

Towards these ends, recent experiments by Pinkerton and
co-workers\cite{pinkerton05, meisner06,pinkerton06} and Aoki
\textit{et al.}\cite{aoki05,nakamori06} on the H-storage properties of
the quaternary Li-B-N-H system are noteworthy.  By mixing lithium
borohydride (\libh) and lithium amide (\amide), both groups have
reported the formation of a new hydride phase, which when heated above
$\sim$520~K released approximately 10 wt.\% hydrogen.  While the
reported desorption temperature is somewhat too high, and
reversibility has not yet been demonstrated, improvement in these
areas may be possible (and has been partially
demonstrated\cite{pinkerton06}) through the addition of catalytic
dopants\cite{bogdanovic97, pinkerton06} or via novel synthetic
routes,\cite{gutowska05} suggesting that further study of this system
is warranted.

The composition of the new Li-B-N-H phase was preliminarily
identified\cite{pinkerton05} as ``Li$_3$BN$_2$H$_8$,'' but subsequent
experiments based on single crystal X-ray
diffraction,\cite{filinchuk06} and, synchrotron X-ray and neutron
powder diffraction,\cite{chater06} identified its true stoichiometry
as \foo. Combined, these diffraction experiments identified three
similar crystal structures, all sharing the bcc space group $I2_13$
with lattice constant $a$ = 10.66--10.68~\AA, but with somewhat
different internal coordinates.  The discrepancy in atomic positions
results in N-H bond lengths that differ between structures by as much
as 0.2~\AA, with smaller N-H lengths reported for the X-ray
structures. It has been suggested that the discrepancy may be an
artifact of X-ray interactions with strongly polarized electron clouds
along the N-H bond direction.\cite{filinchuk06}
 
A careful characterization of a hydride's crystal structure is
desirable because it enables an independent, \textit{ab initio}
assessment of the thermodynamics of hydrogen desorption/absorption
reactions.  Such an assessment is of value because it can clarify the
thermodynamics when calorimetry measurements yield ambiguous results,
such as in situations where multiple reactions occur simultaneously
(see below).  A key thermodynamic property which determines the
suitability of a hydride for H-storage applications is the strength of
the hydrogen-host bond.  The bond strength is quantified via the
change in enthalpy $(\Delta H)$ occurring during hydrogen
uptake/release.  In mobile FC applications, for example, it is
desirable that the desorption reaction be \textit{endothermic}
$(\Delta H > 0)$ with $\Delta H \approx$ 20--50 kJ/(mol \htwo).  Neglecting
kinetics, an enthalpy in this range would allow for H-desorption at
temperatures compatible with the waste heat from a FC, and
permit on-board recharging under reasonable \htwo\ pressure.

Unfortunately, calorimetric measurements of H-desorption in \foo\ have
been difficult to interpret. (See Refs.~\onlinecite{pinkerton06} and
\onlinecite{nakamori06}, and the supplementary information
accompanying Ref.~\onlinecite{pinkerton05}.)  Because \foo\ melts at
$\sim$460~K, approximately 60 degrees below the onset of H-desorption,
hydrogen gas evolves from the liquid state.\cite{pinkerton05}
Concurrent with \htwo\ release is the formation of a solid reaction
product, generally involving one or more polymorphs of \libn\ along
with other unidentified phases,\cite{pinkerton05,aoki05,nakamori06}
and desorption of 2-3 mol \% ammonia.\cite{pinkerton05} Measurements
based on differential scanning calorimetry (DSC)\cite{pinkerton05,
  pinkerton06} and differential thermal analysis
(DTA)\cite{nakamori06} suggest that the \textit{net} heat flow during
dehydriding is exothermic $(\Delta H < 0)$.  However, it is unclear
what fraction of the total thermal profile arises from the exothermic
latent heat of \libn\ solidification,\cite{nakamori06} and furthermore
how \htwo\ and \nhthree\ release impacts the calorimeter
response.\cite{pinkerton05} Consequently, it has not been possible to
unambiguously assess the endo- or exothermic nature of H-desorption
alone.  Neglecting kinetic effects, the fact that it has proven very
difficult\cite{pinkerton05} to re-hydride \foo\ suggests desorption is
weakly endothermic, or exothermic. Regarding \foo\ formation, several
studies\cite{pinkerton05,chater06,nakamori06} have reported the
spontaneous formation of \foo\ after mixing \libh\ with \amide, even
at temperatures as low as room temperature, suggesting that the
formation reaction is exothermic.\cite{pinkerton05}

In light of the discrepancy noted above regarding the crystal
structure of \foo, and the additional ambiguity surrounding its
dehydrogenation thermodynamics, we employ density functional theory
(DFT)\cite{hohenberg64,kohn65} calculations in an attempt to clarify
these issues. First, we determine the ground state crystal structure
by performing separate geometry optimization calculations on each of
the experimentally proposed crystal structures.  We find that the
relaxed structures are essentially identical, and have N-H bond
lengths that are consistent with the structure based on neutron
diffraction,\cite{chater06} thus confirming the
conjecture\cite{filinchuk06} that the anomalously short N-H bond
lengths observed in Ref.~\onlinecite{filinchuk06} are an artifact of
X-ray measurements.  Second, we evaluate the finite-temperature
reaction enthalpies and free energies for the formation and
dehydrogenation of \foo.  \foo\ formation (with respect to \amide\ and
\libh) is found to be exothermic, in agreement with experimental
reports of spontaneous \foo\ formation in the
literature.\cite{pinkerton05, chater06, aoki05} For \htwo-desorption
we explore the thermodynamics of seventeen candidate reactions over
the temperature range $T =$ 0--1000~K, as there appears to be some
uncertainty in determining the reaction products
experimentally.\cite{pinkerton05, aoki05} At temperatures where
desorption has been reported\cite{pinkerton05} ($T \approx $
520-630~K), the favored reaction is predicted to be \foo\ $
\rightarrow $ \libn\ + \amide\ + 4~\htwo, which is weakly endothermic.
Moreover, the calculated free energies suggest \foo\ is a metastable
phase that should decompose via one of three temperature-dependent
pathways.

\section{Methodology}

First-principles calculations were performed using a planewave method
based on the PW91 generalized gradient approximation\cite{perdew92} to
density functional theory\cite{hohenberg64,kohn65}
(\textsc{vasp}).\cite{kresse96} The core-valence electron interaction
was treated using Bl\"{o}chl's projector augmented wave (PAW)
method,\cite{blochl94a,kresse99} and k-point sampling was performed on
a dense Monkhorst-Pack\cite{monkhorst76} grid with an energy
convergence of better than 1~meV per supercell.  (In the case of \foo,
a $2\times 2 \times 2$ grid was sufficient to acheive this level of
precision.)  Electronic occupancies were determined via a gaussian
smearing algorithm with a 0.1~eV smearing width. For high-precision
calculation of static zero Kelvin (omitting zero point vibrational
effects) electronic energies and crystal structures we used the
so-called ``hard'' \textsc{vasp} PAW potentials\footnote{More
  specifically, the potentials employed for evaluating zero-Kelvin
  energies and geometries were (in the parlance of the \textsc{VASP}
  database): Li\_sv, B\_h, N\_h, and H\_h, with respective valence
  electron configurations of $1s^2\,2s^1$, $2s^2\,2p^1$, $2s^2\,2p^3$,
  and $1s^1$.} with a planewave cutoff of 875~eV, and a geometry
relaxation tolerance of better than 0.02~eV/\AA.  Internal atomic
positions and external cell shape/volume were optimized
simultaneously.

Finite-temperature thermodynamics were evaluated within the harmonic
approximation.\cite{wallace72, rickman02} Vibrational frequencies
$(\omega_i)$ were extracted by diagonalizing a dynamical matrix whose
elements were determined via the so-called direct method:\cite{wei92}
the forces generated by series of symmetry inequivalent atomic
displacements about the equilibrium geometry ($\pm 0.02$~\AA, $\pm
0.04$~\AA) were fit to cubic splines in order to extract the force
constants.  Because large supercells are desirable to minimize
finite-size effects, and with the exception of the molecular species
(\htwo, \ntwo, \nhthree) where the number of atoms per supercell is
small, the dynamical matrix calculations on the solid-state phases
were preformed using a softer set of PAW potentials\footnote{The
  geometry of each structure was re-optimized using the softer PAW set
  before the dynamical matrix calculations were performed.} and
planewave cutoff energies ranging from 400--500~eV.\footnote{Our
  somewhat non-standard approach of using softer potentials to
  determine the vibrational spectra of the solid phases, but hard
  (semi-core) potentials for the remaining properties (zero-Kelvin
  energetics, ground state geometries, and molecular vibrational
  spectra,) is based on three considerations: First, the softer PAW
  set allows for more efficient calculations with a reduced planewave
  cutoff energy, thus permitting larger supercells and, consequently,
  smaller finite-size effects in the phonon spectrum.  Second, our
  calculated formation energies for LiH, \amide, and \imide\ at $T =
  300$~K agree with the hard PAW results in Ref.~\onlinecite{herbst05}
  to within 1~kJ/mol f.u.  Third, testing revealed that the hard
  potentials yielded molecular vibrational frequencies which were in
  slightly better agreement with experimental data than those obtained
  with the softer PAW set. For example, for \ntwo, where the
  differences between calculated vibrational frequencies are largest,
  we find 2338 (2452)~cm$^{-1}$ from the hard (soft) PAW sets,
  compared with an experimental value of 2359~cm$^{-1}$.\cite{crc2005}
  However, this slight discrepancy in frequencies has only a small
  impact on vibrational thermodynamic properties---for example, ZPEs
  calculated with both PAW sets differ by 0.7 kJ/mol f.u.\ at most.
  It therefore seems likely that an approach where \textit{all}
  vibrational properties (solids and molecules) were evaluated using
  the softer potentials would also yield reasonably precise results.
  We thus conclude that our approach of mixing potentials gives
  results comparable in precision to those obtained with the hard PAW
  set alone, but without the associated computational expense.}  A
comparison of calculated structural parameters and formation energies
with existing experimental and theoretical data is presented in the
following section.

Once the normal-mode frequencies have been determined,
finite-temperature energetics can be obtained by enthalpic
($H_\text{vib}$) and entropic (\svib) additions to the static
electronic energies.  Within the harmonic approximation these
contributions are given by:\cite{wallace72}
\begin{eqnarray}
\label{hvib}
H_\text{vib}(T) &=& \sum_i\frac{1}{2} \hbar\omega_i + \hbar\omega_i\left[\exp(\frac{\hbar \omega_i}{k_B T}) -1\right]^{-1}\\
\label{svib}
S_\text{vib}(T) &=& k_B \sum_i \frac{\hbar \omega_i/k_B T}{\exp\left(\frac{\hbar \omega_i}{k_B T}\right ) -1}\nonumber \\&& -\ln\left[1 - \exp(\frac{-\hbar \omega_i}{k_B T})\right],
\end{eqnarray}
where the sums run over vibrational frequencies ($3N-3$ frequencies
for solids and non-linear molecules, $3N-5$ frequencies for linear
molecules), $k_B$ is the Boltzmann factor, and $T$ is the absolute
temperature.  For the linear (non-linear) molecules an additional
$\frac{7}{2} k_BT$ $(4 k_BT)$ term is added to Eq.~\ref{hvib} to
account for translational, rotational, and $pV$ degrees of freedom.
The zero point energy (ZPE) can be recovered from Eq.~\ref{hvib} in
the limit \hvib$(T = 0)$.  The enthalpy and free
energy of a phase can therefore be expressed as:
\begin{eqnarray}
H(T) &=& E + H_\text{vib}(T) \label{h}\\
G(T) &=& H(T) - S(T)T \label{g}, 
\end{eqnarray}
where $E$ is the static electronic energy of the crystal/molecule in
its ground-state geometry and $S$ represents either the standard
tabulated\cite{janaf98} entropy of a given molecular species in the
gas phase at $p = 1$ bar [$S_0^\text{T=300K} =$ 130.858 (\htwo),
191.789 (\ntwo), and 192.995 J K$^{-1}$ mol$^{-1}$ (\nhthree)], or the
vibrational entropy \svib\ of a solid state phase.

\section{Crystal Structure}
The crystal structure of \foo\ has been studied by two
groups.\cite{filinchuk06, chater06} Filinchuk and
co-workers\cite{filinchuk06} used single-crystal X-ray diffraction to
identify the structures of the two largest crystal domains obtained
after remelting a 2:1 mixture of \amide\ and \libh.  Both domains
exhibited a bcc structure (space group $I2_13$) with lattice constants
of 10.67-10.68~\AA.  Chater \textit{et al.}\cite{chater06} used a
combination of high-resolution synchrotron X-ray and neutron
diffraction to examine powder samples prepared from a wide range of
\amide:\libh\ compositions.  They determined that although the most
likely stoichiometric composition was 3:1, the \foo\ structure was
able to accommodate a wide range of stoichiometries.  Their best-fit
crystal structure had external cell parameters similar to those
reported in Ref.~\onlinecite{filinchuk06}: bcc (space group $I2_13$)
with $a = 10.66$~\AA.  Despite the good agreement in external
geometries, the structures obtained by these two studies differ
markedly in their internal coordinates. Most notably, the
single-crystal X-ray diffraction study\cite{filinchuk06} found
anomalously short N-H bond lengths of 0.83-0.86~\AA, while neutron
scattering\cite{chater06} gave lengths of 0.98-1.04~\AA.

\begin{table}
  \caption{Calculated structural parameters compared with experimental data. Bond lengths ($d$) and crystal lattice constants ($a$, $b$, $c$) are given in \AA, bond angles ($\measuredangle$) and angles between lattice vectors ($\alpha$, $\beta$) are listed in degrees.  Space group information for the solid phases is listed in parentheses.}
\label{structure}
\begin{ruledtabular}
\begin{tabular}{lcrr}
System & Parameter & Calculated& Experiment\\ \hline
\htwo & $d$(H-H) & 0.749 & 0.741\cite{crc2005}\\
\ntwo & $d$(N-N) & 1.101 & 1.098\cite{crc2005}\\
\nhthree & $d$(N-H)& 1.021 & 1.012\cite{crc2005}\\
         & $\measuredangle$(H-N-H) & 106.4 & 106.7\cite{crc2005}\\
$\alpha$-B ($R\bar{3}m$) & $a$ & 5.05 & 5.06\cite{switendick91}\\
&          $\alpha$ & 58.1 & 58.1\cite{switendick91}\\
BN ($F\bar{4}3m$) & $a$ & 3.62 & 3.62\cite{wentorf57} \\
Li ($Im\bar{3}m$)   & $a$      &3.44&   3.51\cite{villars91}\\
LiH ($Fm\bar{3}m$) & $a$ & 4.01 & 4.07\cite{vidal86} \\
Li$_3$N ($P\bar{3}m1$) & $a$ & 3.64 &3.61\footnotemark[1] \\
&  $c$ & 3.88 & 3.85\footnotemark[1] \\
\amide\ ($I\bar{4}$) & $a$ &5.02 & 5.04\cite{jacobs72}\\
       & $c$ &10.28 & 10.28\cite{jacobs72}\\
\imide\ ($Pnma$) & $a$ &7.75 & 7.73\footnotemark[2] \\
       & $b$ &3.61 & 3.60\footnotemark[2]\\
       & $c$ &4.88 & 4.87\footnotemark[2]\\
$\alpha$-\libn\ ($P4_2/mnm$) & $a$ & 4.67 & 4.64\cite{cenzual91,yamane87}\\
       & $c$ & 5.22 & 5.26\cite{cenzual91,yamane87} \\
$\beta$-\libn\ ($P2_1/c$) & $a$ & 5.15 & 5.15\cite{yamane86}\\
       & $b$ & 7.08 & 7.08\cite{yamane86}\\
       & $c$ & 6.77 & 6.79\cite{yamane86}\\
       & $\beta$ & 112.7 & 113.0\cite{yamane86}\\
\libn\ ($I4_1/amd$) & $a$ & 6.63 & 6.60\cite{pinkerton06a} \\
       & $c$ & 10.35 & 10.35\cite{pinkerton06a}\\
\libh\ ($Pnma$)  & $a$ & 7.25 & 7.18\cite{soulie02}\\
       & $b$ & 4.38 & 4.44\cite{soulie02}\\
       & $c$ & 6.63 & 6.80\cite{soulie02}\\
\foo\ ($I2_13$)  & $a$ & 10.60 & 10.67-10.68\cite{filinchuk06}\\
       &     &       & 10.66\cite{chater06}
\end{tabular}
\footnotetext[1]{Theoretically predicted structure from
  Ref.~\onlinecite{miwa05}.  In agreement with
  Ref.~\onlinecite{miwa05}, we find the room-temperature
  experimental $P6/mmm$ structure to be unstable at low temperature,
  with a weak soft mode of 66$i$ cm$^{-1}$.}
\footnotetext[2]{Theoretically predicted structure from
  Ref.~\onlinecite{magyari-kope06}.}
\end{ruledtabular}
\end{table}

The external cell parameters obtained upon relaxing each of the three
\foo\ experimental structures (as well as the structures of the other
phases used in our subsequent discussion of reaction thermodynamics)
are summarized in Table~\ref{structure}.  The resulting structures are
bcc and share the same lattice constant, $a = 10.60$~\AA, in good
agreement with both diffraction studies.

\begin{table*}
\caption{Calculated relaxed internal atomic positions of \foo\ compared with experimental measurements from Refs.~\onlinecite{filinchuk06, chater06}.}
\label{geometry}
\begin{ruledtabular}
\begin{tabular}{crrrrrr|c}
& \multicolumn{3}{c}{Calculated} & \multicolumn{3}{c}{Experiment} & \\\cline{2-4}\cline{5-7}
Atom & \multicolumn{1}{c}{\textit{x}} & \multicolumn{1}{c}{\textit{y}} & \multicolumn{1}{c}{\textit{z}} & \multicolumn{1}{c}{\textit{x}} & \multicolumn{1}{c}{\textit{y}} & \multicolumn{1}{c}{\textit{z}} & \\\hline
N       &0.117  &0.362  &0.402  &0.115  &0.359  &0.405& \\
B       &0.113  &0.113  &0.113  &0.114  &0.114  &0.114&\\
Li1     &0.281  &0.000  &0.250  &0.287  &0.000  &0.250&\\
Li2     &0.519  &0.000  &0.250  &0.524  &0.000  &0.250&Domain 1,\\
Li3     &0.484  &0.484  &0.484  &0.483  &0.483  &0.483&\\
H1      &0.195  &0.306  &0.393  &0.175  &0.310  &0.397&Ref.~\onlinecite{filinchuk06}\\
H2      &0.118  &0.415  &0.321  &0.115  &0.405  &0.339&\\
H3      &0.003  &0.118  &0.147  &0.012  &0.119  &0.146&\\
H4      &0.180  &0.180  &0.180  &0.172  &0.172  &0.172&\\\hline
N       &0.117  &0.361  &0.402  &0.115  &0.359  &0.405& \\
B       &0.113  &0.113  &0.113  &0.114  &0.114  &0.114&\\
Li1     &0.281  &0.000  &0.250  &0.286  &0.000  &0.250&\\
Li2     &0.519  &0.000  &0.250  &0.523  &0.000  &0.250&Domain 2,\\
Li3     &0.484  &0.484  &0.484  &0.484  &0.484  &0.484&\\
H1      &0.195  &0.305  &0.394  &0.176  &0.310  &0.399&Ref.~\onlinecite{filinchuk06}\\
H2      &0.118  &0.415  &0.321  &0.114  &0.402  &0.338&\\
H3      &0.003  &0.117  &0.148  &0.015  &0.116  &0.149&\\
H4      &0.180  &0.180  &0.180  &0.174  &0.174  &0.174&\\\hline
N       &-0.152 &0.112  &0.367  &-0.156 &0.110  &0.367& \\
B       &0.137  &0.137  &0.137  &0.135  &0.135  &0.135&\\
Li1     &0.250  &-0.030 &0.000  &0.250  &-0.043 &0.000&\\
Li2     &-0.268 &0.000  &0.250  &-0.262 &0.000  &0.250&\\
Li3     &0.266  &0.266  &0.266  &0.272  &0.272  &0.272&Ref.~\onlinecite{chater06}\\
H1      &0.071  &0.071  &0.071  &0.072  &0.072  &0.072&\\
H2      &0.103  &0.133  &0.247  &0.097  &0.136  &0.248&\\
H3      &-0.143 &0.055  &0.445  &-0.153 &0.053  &0.439&\\
H4      &-0.071 &0.165  &0.368  &-0.077 &0.167  &0.377&\\
\end{tabular}
\end{ruledtabular}
\end{table*}

For each of the three relaxed structures, Table~\ref{geometry}
compares the calculated internal atomic coordinates with the
corresponding experimental values.  We note first that the average
absolute deviation $(\delta)$ between theory and experiment is
smallest for the neutron structure,\cite{chater06} $\delta = 3.6\times
10^{-3}$, whereas for the X-ray structure\cite{filinchuk06} $\delta =
4.4\times 10^{-3}$ and $\delta = 4.2\times 10^{-3}$ for domains 1 and
2, respectively.  The largest discrepancy between theory and
experiment is in the positions of the N-bonded hydrogen atoms, H1 \&
H2, in the X-ray structure.  (Note that the labeling convention for
hydrogens is different in the neutron structure: there H1 \& H2 refer
to B-bonded hydrogens.)  Secondly, while there appear to be large
differences in the internal coordinates measured by
Refs.~\onlinecite{filinchuk06} and \onlinecite{chater06}
(Table~\ref{geometry}), after relaxation these three structures may be
mapped onto one another via a series of rigid body rotations and
translations.  We conclude that all three experimental structures
relax to essentially the same structure.

\begin{table*}
  \caption{Comparison of calculated atomic distances and angles with values determined from single-crystal X-ray diffraction (Ref.~\onlinecite{filinchuk06}) and synchrotron X-ray and neutron powder diffraction (Ref.~\onlinecite{chater06}).  Lengths ($d$) are given in \AA, angles ($\measuredangle$) are in degrees.}
\label{geometry2}
\begin{ruledtabular}
\begin{tabular}{lrrrrrr}
  & \multicolumn{2}{c}{Ref.~\onlinecite{filinchuk06}, domain 1} & \multicolumn{2}{c}{Ref.~\onlinecite{filinchuk06}, domain 2} & \multicolumn{2}{c}{Ref.~\onlinecite{chater06}} \\\cline{2-3}\cline{4-5}\cline{6-7}
  Parameter & Calculated & Experiment & Calculated & Experiment & Calculated & Experiment\\\hline
  \multicolumn{7}{c}{Intramolecular Values}\\
  $d$(N-H) (\textit{i})& 1.027 & 0.83(2) & 1.027 & 0.84(2) & 1.027 & 0.983 \\
  $d$(N-H) (\textit{ii})& 1.027 & 0.859   & 1.026 & 0.852   & 1.027 & 1.042 \\
  $d$(B-H) (\textit{i})& 1.223 & 1.08(3) & 1.225 & 1.11(3) & 1.224 & 1.169 \\
  $d$(B-H) (\textit{ii})$\times 3$ & 1.225 & 1.137 & 1.225 & 1.121 & 1.223 & 1.270\\
  $\measuredangle$(H-N-H) & 103.5 & 106.2 & 103.5 & 105.9 & 103.5 & 104.9\\
  $\measuredangle$(H-B-H)$\times 3$ (\textit{i}) & 108.0 & 108.9 & 107.9 & 109.2 & 107.9 & 107.6\\
  $\measuredangle$(H-B-H)$\times 3$ (\textit{ii}) & 110.9 & 110.1 & 111.0 & 109.8 & 111.0 & 111.3 \\
  \multicolumn{7}{c}{Intermolecular Values}\\
  $d$(N-Li) (\textit{i}) & 2.071 & 2.069 & 2.069 & 2.068 & 2.070 & 2.033\\
  $d$(N-Li) (\textit{ii}) & 2.116 & 2.115 & 2.111 & 2.113 & 2.110& 2.053 \\
  $d$(N-Li) (\textit{iii}) & 2.138 & 2.117 & 2.138 & 2.119 & 2.142& 2.094\\
  $d$(N-Li) (\textit{iv}) & 2.164 & 2.157 & 2.164 & 2.157 & 2.169& 2.209\\
  $d$(B-Li) (\textit{i}) & 2.371 & 2.410 & 2.375 & 2.406 & 2.368 & 2.527\\
  $d$(B-Li) (\textit{ii})$\times 3$ & 2.595 & 2.651 & 2.592 & 2.642 & 2.589 & 2.681\\
\end{tabular}
\end{ruledtabular}
\end{table*}

A comparison of calculated bond lengths, bond angles, and
intermolecular distances is given in Table~\ref{geometry2}.  Overall,
the relaxed structures agree best with the \textit{intra}molecular
lengths (i.e., within an NH$_2$ or BH$_4$ fragment) determined using
neutron diffraction; conversely, the \textit{inter}molecular distances
from X-ray are closest to those in the relaxed theoretical
structures. As for the intramolecular structure, as noted above, the
major discrepancy between the experimental structures lies in the two
N-H bond lengths, $d$(N-H), in the NH$_2$ fragment.  The calculated
length of 1.027~\AA\ is in better agreement with the longer bond
lengths predicted by neutron diffraction of 0.983 and 1.042~\AA.
Similarly, the X-ray data also underestimates B-H bond lengths in the
BH$_4$ units relative to our DFT calculations and neutron data.
Calculated bond angles agree best with those from the neutron
structure.

\section{Reaction Energetics}
In order to examine the finite-temperature thermodynamics of reactions
involving \foo, it was first necessary to calculate the ground-state
structures of the phases that participate in those reactions.  A
summary of the compounds used, their crystal structures, and a
comparison of the calculated structures to experimental data is
presented in Table~\ref{structure}.  In general the agreement between
experiment and theory is very good.

Some of the compounds listed in Table~\ref{structure} are known to
have several polymorphs.  For example, at least three polymorphs have
been reported for \libn,\cite{yamane86,yamane87,wentorf61,devries69}
including low-temperature tetragonal ($\alpha$),\cite{yamane87}
high-temperature monoclinic ($\beta$),\cite{yamane86} and
high-pressure body-centered tetragonal (bct)
\cite{wentorf61,devries69, pinkerton06a} phases.  All three of these
phases have been observed as dehydrogenation products for
\foo,\cite{pinkerton05, nakamori06} and we have performed calculations
on each phase to determine its relative stability as a function of
temperature. The $\beta$-\libn\ phase was found to have the lowest
zero-Kelvin static energy, $\approx$ 2~kJ/(mol f.u.) lower than that
of either the bct or $\alpha$ phases.  The latter two phases are
degenerate in energy to within 0.1~kJ/(mol f.u.).  Our zero-Kelvin
energetics and structures agree well with recent calculations reported
by Pinkerton and Herbst.\cite{pinkerton06a}

Vibrational contributions to the energies of each \libn\ polymorph
were evaluated within the harmonic approximation.  For $\alpha$-\libn\
a careful evaluation of the vibrational spectra yielded
doubly-degenerate imaginary modes at 81$i$~cm$^{-1}$, indicating that
this phase is unstable at low temperatures. A search for alternative
low-energy $\alpha$-\libn\ structures was performed using molecular
dynamics (md) on an enlarged $(2\times2\times2)$ $\alpha$-\libn\
supercell at $T =$ 423~K for 10~ps.  At 1~ps intervals the current md
configuration was stored, relaxed, and the symmetry of the resulting
optimized structure was determined.  Four unique crystal structures
were identified, with space groups (numbers): $P2_12_12_1$ (19),
$Pccn$ (56), $Pmmn$ (59), and $Pnma$ (62).  These structures were then
relaxed within their symmetry constraints, and were found to have
energies $\sim$1~kJ/(mol f.u.)  lower than the original
$\alpha$-\libn\ structure [which is still 1~kJ/(mol f.u.)  higher in
energy than $\beta$-\libn]. In light of these results we suggest that
the crystal structure of $\alpha$-\libn\ be experimentally
re-assessed, and we exclude this phase from our evaluation of \foo\
dehydrogenation reactions.  Further information regarding our proposed
structure of $\alpha$-\libn\ can be found
elsewhere.\cite{siegel-unpub}

\begin{figure}
\centering \includegraphics[width=\linewidth]{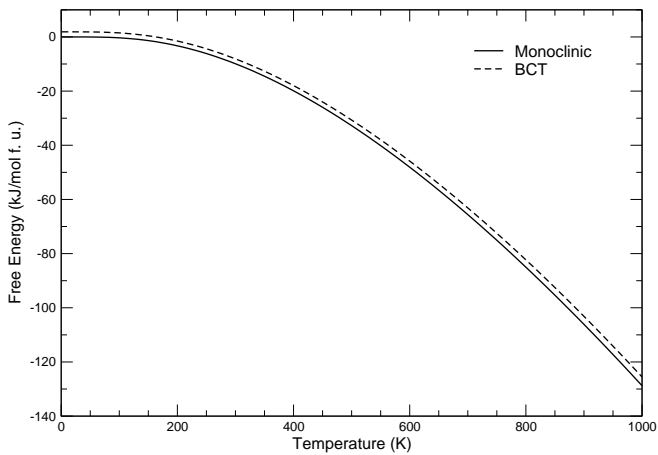}
\caption{Calculated Gibbs free energies (in kJ/mol f.u.) of monoclinic
  ($\beta$) and body-centered tetragonal (BCT) \libn\ as a function of
  temperature.  The energy zero is set to the static 0~K energy of
  $\beta$-\libn. }
\label{fig2}
\end{figure}

Unlike $\alpha$-\libn, no imaginary modes were observed in the
vibrational spectra of $\beta$ or bct \libn.  Using
Eqs.~\ref{hvib}-\ref{svib}, Fig.~\ref{fig2} plots the Gibbs free
energies (Eq.~\ref{g}) of these phases as a function of
temperature.  Due to their similar vibrational contributions
(Table~\ref{energies}), the free energies of the $\beta$ and bct
phases exhibit nearly identical temperature dependence.  Consequently,
the small $\sim$2 kJ/(mol f.u.)  difference in static energies noted
above is sufficient to favor the monoclinic $\beta$ phase as the
ground state structure, and the bct-$\beta$ energy difference remains
roughly constant for $T < 1000$~K.  Based on the vibrational
instability observed in $\alpha$-\libn, and the higher free energy of
bct \libn\ (relative to $\beta$-\libn), our subsequent thermodynamic
analyses are performed assuming \libn\ resides in the the low-energy
$\beta$ structure.

\begin{table}
  \caption{Calculated vibrational contributions (at $T = 300K$) to the free energies of phases used in this study.  ZPE refers to the zero point energy, given by \hvib($T=0$) (Eq.~\ref{hvib}); \evib\ = \hvib\ $-$ ZPE (omitting the $k_BT$ molecular terms); \svib\ is the vibrational entropy, Eq.~\ref{svib}. Units are kJ~mol$^{-1}$ for ZPE and \evib, and J mol$^{-1}$ K$^{-1}$ for \svib.}
\label{energies}
\begin{ruledtabular}
\begin{tabular}{lrrrr}
System & ZPE & ZPE other & \evib$^\text{T=300K}$ & \svib$^\text{T=300K}$ \\\hline 
\htwo    & 25.7  & 25.7\cite{frankcombe06}, 26.1\cite{miwa04}, &$\approx 0$ & $\approx 0$\\
 & & 26.3\cite{herbst05}& & \\
\ntwo    & 14.0  & 14.4\cite{herbst05}&  $\approx 0$  & $\approx 0$ \\
\nhthree & 87.8  & 89.7\cite{northrup97}&  0.1 & 0.4 \\
Li       & 3.9   & 3.9\cite{miwa04}&  4.3  & 27.7 \\
B        & 12.5  & 12.2\cite{miwa04}&  1.1  & 5.3\\
BN       & 29.9  & 30.9\cite{kern99} &      &    \\
LiH      & 21.7  & 21.4,\cite{miwa04} 21.8\cite{magyari-kope06}&  3.5  & 17.7\\
Li$_3$N  & 28.0  & 28.6\cite{miwa05} & & \\
\amide   & 69.0  & 69.1,\cite{miwa05} 69.5,\cite{magyari-kope06} &  8.9  & 51.6 \\
         &       & 69.8\cite{herbst05} & & \\
\imide   & 46.7  & 46.7\cite{herbst05}, 47.1\cite{magyari-kope06}&  9.6  & 52.9\\
\libh    & 107.1 & 103.0,\cite{miwa04} 106.5,\cite{frankcombe06}&  10.8 &63.6\\
& &  108.1\cite{lodziana04} & & \\
\libn    & 52.2  & &  15.3 & 83.8\\
\foo     & 314.4 & &  37.0 & 213.4\\
\end{tabular}
\end{ruledtabular}
\end{table}

\begin{table}
  \caption{Comparison of calculated formation energies (with respect to the elements, in kJ/mol f.u.) with existing experimental and theoretical data. $\Delta E$ refers to differences in static 0~K energies, formation enthalpies ($\Delta H$) and free energies ($\Delta G$) are based on Eqns.~\ref{h} and \ref{g}, respectively. Experimental data is from Refs.~\onlinecite{janaf98} and \onlinecite{chen02}, theoretical data is from a set of recent DFT calculations (Refs.~\onlinecite{herbst05, miwa04, pinkerton06a}).}
\label{form}
\begin{ruledtabular}
\begin{tabular}{lrrr}
System & $\Delta E$ & $\Delta H^\text{T=300K}$ & $\Delta G^\text{T=300K}$ \\\hline
\nhthree &$-$98.0   & $-$63.2 & $-$33.6  \\
         &   & $-$45.9\cite{janaf98} & $-$16.2\cite{janaf98}  \\
BN       &$-$233.8   &$-$227.4 & $-$198.9\\
         &   &$-$250.9\cite{janaf98} & $-$224.9\cite{janaf98}\\
LiH      &$-$83.6   &$-$83.8  & $-$61.2 \\
         &$-$81\cite{miwa04}   &$-$90.6\cite{janaf98}  & $-$68.3\cite{janaf98}\\
         &$-$83.9\cite{herbst05}   &$-$84.6\cite{herbst05}  &  \\
Li$_3$N  &$-$148.7   &$-$145.9 & $-$109.8\\
         &   &$-$164.6\cite{janaf98} &$-$128.4\cite{janaf98}\\
\amide   &$-$196.6   &$-$172.6 &$-$111.7  \\
         &$-$196.5\cite{herbst05}   &$-$176\cite{chen02}   & \\
         &   &$-$173.1\cite{herbst05} & \\
\imide   &$-$196.0  &$-$184.5 & $-$135.3  \\
         &$-$194.0\cite{herbst05}   &$-$222\cite{chen02}   & \\
         &   &$-$184.1\cite{herbst05} & \\
\libh    &$-$205.9   &$-$178.6 &$-$109.2  \\
         &$-$194\cite{miwa04}   &$-$190.5\cite{janaf98} & $-$124.3\cite{janaf98}\\ 
\libn    &$-$497.4   &$-$490.6 &         $-$433.6 \\
         &$-$497\cite{pinkerton06a} & & \\
\end{tabular}
\end{ruledtabular}
\end{table}

The zero-point, enthalpic, and entropic contributions to the free
energy of the various phases are summarized in Table~\ref{energies}.
As a further test of our computational methodology, there we also
compare our calculated ZPE to other values reported in the literature,
and find good agreement.  With the exception of Li, vibrational
effects are found to contribute substantially to the free energies of
these systems. For example, in \foo\ the ZPE exceeds 300 kJ/mol.

A comparison of formation energies for the compounds used in this
study (relative to the elements in their standard states) is presented
in Table~\ref{form}.  Three reports\cite{miwa04, herbst05,
  pinkerton06a} based on DFT calculations have also recently evaluated
formation energies for some of these compounds, and the present
calculations are in good agreement with their findings. Furthermore,
the agreement with experimental enthalpies is reasonable (generally on
the order of 10\%), which is representative of the accuracy obtained
via density functional methods for reactions involving molecular
species.\cite{curtiss97} As expected, the largest percentage
discrepancy is for the ammonia molecule.\footnote{The discrepancy in
  calculated vs.\ experimental enthalpy for \imide\ has been noted in
  Ref.~\onlinecite{herbst05}, and it has been suggested that the
  experimental measurement be revisited in light of the good agreement
  for the other Li-containing compounds: LiH, \amide, and \libh.  To
  our knowledge the formation enthalpy of \imide\ reported in
  Ref.~\onlinecite{chen02} has not been independently confirmed.}  Two
general trends in the data are evident: (\textit{i}) With the
exception of ammonia, the DFT enthalpies ($\Delta H^\text{T=300K}$)
are more positive than the experimental values, and (\textit{ii}) the
inclusion of dynamical contributions ($\Delta E \rightarrow \Delta
H^\text{T=300K}$) results in more positive formation energies.

\begin{table*}
  \caption{Formation energies of \foo\ [in kJ/(mol \foo)].   $\Delta E$ corresponds to the static zero Kelvin DFT energies; $\Delta H^\text{T = 0K}$ adds the zero point energies to $\Delta E$; $\Delta H^\text{T= 300K}$ (Eq.~\ref{h}) adds finite-temperature vibrational (and molecular rotational+translational+$pV$) energies to $\Delta H^\text{T=0K}$; $\Delta G^\text{T=300K}$ (Eq.~\ref{g}) includes all of the contributions to $\Delta H^\text{T=300K}$ and adds vibrational (solids) or tabulated (molecular) entropies.}
\label{formation}
\begin{ruledtabular}
\begin{tabular}{clrrrr}
Rxn.\ No. & \multicolumn{1}{c}{Reaction} & $\Delta E$ &$\Delta H^\text{T=0K}$ & $\Delta H^\text{T=300K}$ & $\Delta G^\text{T=300K}$ \\ \hline
(a) & 4 Li + B + $\frac{3}{2}$\ntwo\ + 5\htwo $\,\rightarrow\,$ \foo & $-$806.9 & $-$670.2 & $-$708.1 &$-$454.6\\
(b) & 3 \amide\ + \libh $\,\rightarrow\,$  \foo &  $-$11.3 & $-$11.2 & $-$11.8 & $-$10.2\\
\end{tabular}
\end{ruledtabular}
\end{table*}

Using the energetic contributions from Table~\ref{energies}, in
Table~\ref{formation} we evaluate the thermodynamics for two \foo\
formation reactions.  Reaction (a) considers formation from the
elemental phases in their standard states, and reaction (b)
corresponds to the synthesis route employed in the
literature\cite{pinkerton05, chater06, aoki05} involving a mixture of
\amide\ and \libh.  (We consider only the
stoichiometric\cite{chater06} 3:1 ratio of \amide\ to
\libh.\footnote{Due to the computational difficulty associated with
  treating a non-stoichiometric phase, our calculations are restricted
  to the stoichiometric composition ``\foo,'' whereas most experiments
  have focused on 2:1 \amide:\libh\ off-stoichiometric
  ``Li$_3$BN$_2$H$_8$'' mixtures, which have been found to
  minimize\cite{meisner06} \nhthree\ release.}) The energetics of each
reaction are are decomposed into a sequence having increasingly more
physical contributions to the reaction thermodynamics, $\Delta E
\rightarrow \Delta H^\text{T=0K} \rightarrow \Delta H^\text{T=300K}
\rightarrow \Delta G^\text{T=300K}$, which allows us to gauge the
importance of these contributions in comparison to the common practice
of evaluating only static zero-Kelvin energetics.  [$\Delta E$ refers
to differences in the static zero-Kelvin energies ($E$ in
Eq.~\ref{h}), and $\Delta H^\text{T=0 K}$ adds ZPE contributions to
$\Delta E$.]  Both formation reactions in Table~\ref{formation} are
predicted to be exothermic ($\Delta H < 0$), in agreement with reports
of spontaneous \foo\ formation from the literature.\cite{pinkerton05,
  aoki05, chater06} As expected, formation from the elements vs.\ from
the hydride mixture is a substantially more exothermic process.

\begin{table*}
  \caption{Calculated reaction energies [$\Delta E$ in kJ/(mol \htwo)], enthalpies [$\Delta H$ in kJ/(mol \htwo)], and free energies [$\Delta G$ in kJ/mol products]  for candidate \foo\ dehydrogenation reactions.}
\label{reactions}
\begin{ruledtabular}
\begin{tabular}{clrrrrrr}
Rxn.\ No.& \multicolumn{1}{c}{Reaction} & $\Delta E$ &$\Delta H^\text{T=0K}$ & $\Delta H^\text{T=300K}$& $\Delta H^\text{T=460K}$ & $\Delta G^\text{T=300K}$& $\Delta G^\text{T=460K}$ \\ \hline
(c) &\foo $\,\rightarrow\,$  \libn\ + \amide\ + 4\htwo\ & 28.2 & 5.7 & 11.2 & 12.5 & $-$88.8 & $-$161.3 \\
(d) &\foo $\,\rightarrow\,$  \libn\ + $\frac{1}{2}$\imide\ + $\frac{1}{2}$\nhthree + 4\htwo\ & 40.6& 17.7&  23.4 & 24.3 & $-$61.4 & $-$145.2\\
(e) &\foo $\,\rightarrow\,$  \libn\ + LiH + \nhthree\ + 3\htwo\ & 42.6 & 17.5 & 23.5 & 24.0 &$-$71.6 & $-$148.0\\
(f) &\foo $\,\rightarrow\,$  \libn\ + LiH + $\frac{1}{2}$\ntwo\ + $\frac{9}{2}$\htwo &  50.2 & 24.1 &29.7 &30.8 &$-$38.2 & $-$131.0\\
(g) &\foo $\,\rightarrow\,$  \libn\ + Li + $\frac{1}{2}$\ntwo + 5\htwo\ & 61.9 & 37.4 & 43.5 & 44.6& 23.0 & $-$82.2\\
(h) &\foo $\,\rightarrow\,$  \libn\ + Li + \nhthree\ + $\frac{7}{2}$\htwo\ & 60.4 & 37.5& 44.1 & 44.7 &$-$10.5 & $-$99.2\\
(i)&\foo $\,\rightarrow\,$  2\amide\ + 2LiH + BN  + 2\htwo             & 6.4 &$-$19.4 &$-$16.1 &$-$15.4 &$-$90.1& $-$121.2\\
(j)&\foo $\,\rightarrow\,$  \amide\ + \lin\ +  BN + 4\htwo             & 56.9&35.8 &40.5    &41.3 &34.1      & $-$34.9 \\
(k)&\foo $\,\rightarrow\,$  4LiH + BN + \ntwo\ + 3\htwo                & 79.6&44.1 &48.5    &48.8 &11.1      & $-$60.7 \\
(l)&\foo $\,\rightarrow\,$  2\imide\ + BN + 4\htwo                     & 45.3&23.3 &27.9    &28.8 &$-$14.9   & $-$83.3 \\
(m)&\foo $\,\rightarrow\,$  2\amide\ + BN + 2Li + 3\htwo               & 60.0&39.5 &45.1    &46.2 &32.3      & $-$23.6 \\
(n)&\foo $\,\rightarrow\,$  \amide\ + \imide\ + LiH + BN +  3\htwo     & 32.3&9.1 &13.2    &14.1 &$-$52.5   & $-$102.3\\
(o)&\foo $\,\rightarrow\,$  \imide\ + 2LiH + BN + \nhthree\ + 2\htwo   & 55.9&28.5 &32.7    &32.1 &$-$35.4   & $-$88.9 \\
(p)&\foo $\,\rightarrow\,$  \amide\ + 3LiH + BN + \nhthree\ + \htwo    & 27.7&$-$9.1 &$-$6.6  &$-$8.8 &$-$73.0 & $-$107.9\\
(q)&\foo $\,\rightarrow\,$  \lin\ + LiH + BN + \nhthree\ + 3\htwo      & 80.9&57.7 &62.6    &62.4 &51.3      & $-$21.6\\
(r)&\foo $\,\rightarrow\,$  $\frac{1}{2}$\imide\ + \lin\ + BN + $\frac{1}{2}$\nhthree\ + 4\htwo\ & 69.3 & 47.8&52.7 &53.2 &61.5 &$-$18.8 \\
(s)&\foo $\,\rightarrow\,$ \libn\ + $\frac{1}{3}$\lin\ + $\frac{2}{3}$\nhthree\ + 4\htwo &48.7 & 26.0 & 31.7&32.5&$-$35.9 & $-$123.6 \\
\end{tabular}
\end{ruledtabular}
\end{table*}

It is noteworthy that dynamical contributions to the reaction
energies are significant in reaction (a) $(\Delta E \neq \Delta H)$,
but appear to play a minor role in reaction (b) $(\Delta E \approx
\Delta H)$.  By examining the values presented in Table~\ref{energies}
for the phases participating in reaction (b) (\foo, \amide, \& \libh)
we see that these terms are sizable, but that they largely cancel out.
This cancellation effect can be understood by the similarities in
internal bonding shared by these three phases: \foo, with its Li$^+$,
BH$_4^-$, and NH$_2^-$ ions is essentially an amalgam of the
structures present in \amide\ and \libh.

Turning now to the dehydrogenation of \foo, we note first that
although several dehydrogenation products have been observed in
experimental studies,\cite{pinkerton05, meisner06,aoki05,nakamori06}
the precise identity and respective proportions of these phases have
not yet been definitively determined.\cite{pinkerton05,aoki05} For
example, Aoki \textit{et al.} reported unidentified diffraction peaks
after dehydrogenation at $2\theta =$ 25$^\circ$ and 35$^\circ$.  A
technique for predicting possible decomposition pathways would be of
significant value in helping to understand these reactions.  However,
decomposition pathways and products are difficult to predict \textit{a
  priori}. To this end, we have scanned through the thermodynamics of
a large number of candidate dehydrogenation reactions over a wide
temperature range in order to identify the energetically favored
products.\footnotemark[\value{footnote}]

\begin{figure}
\centering \includegraphics[width=\linewidth]{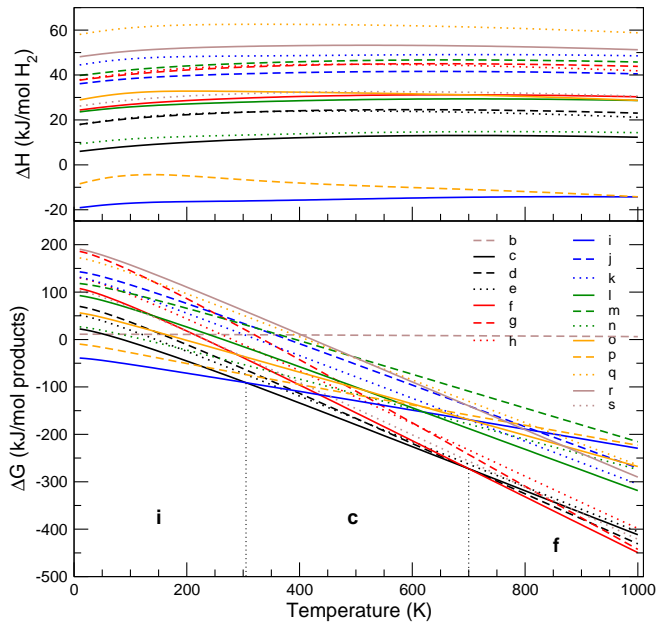}
\caption{(Color online) Calculated enthalpies ($\Delta H$, top panel)
  and free energies ($\Delta G$, bottom panel) of candidate \foo\
  dehydrogenation reactions as a function of temperature at $p = 1$
  bar. The reactions are labeled as in Table~\ref{reactions}.}
\label{fig1}
\end{figure}

Calculated thermodynamics for seventeen candidate \foo\ dehydrogenation
reactions are listed in Table~\ref{reactions}.  In addition to the
room-temperature energetics, there we also present data for $T =
460$~K, which is slightly below the melting point of pure \foo.
Unlike the \foo\ formation reaction, which can proceed at room
temperature,\cite{pinkerton05} dehydrogenation of \foo\ has been
reported only at elevated temperatures.  In the case of pure \foo,
dehydrogenation occurs above $\sim$520~K, about 60 degrees above the
melting temperature.\cite{pinkerton05} However, a significant
reduction in dehydrogenation temperature was observed upon the
addition of a small amount of a Pt/Vulcan carbon
catalyst.\cite{pinkerton06} In this latter case hydrogen release
occurred from the solid phase at $T > 390$~K.  To more completely
assess the dehydrogenation thermodynamics over a range of relevant
temperatures, in Fig.~\ref{fig1} we plot the enthalpies and free
energies of the candidate reactions for $T =$ 0--1000~K.

Out of the seventeen reactions considered, three reactions---(i), (c),
and (f)---emerge as the most-favorable reactions in distinct
temperature regimes (see Fig.~\ref{fig1}, bottom panel).  For low
temperatures reaction (i) is preferred, followed at increasing
temperatures by reaction (c) and reaction (f).  More specifically, the
favored products and their respective temperature ranges of stability
are:
\begin{tabular}{r@{T}lr}
0~K $\leq$ & $\leq$ 300~K: & 2\amide\ + 2LiH + BN + 2\htwo\\
300~K $\leq$ & $\leq$ 700~K: & \libn\ + \amide\ + 4\htwo\\
700~K $\leq$ &: & \libn\ + LiH + $\frac{1}{2}$\ntwo\ + $\frac{9}{2}$\htwo\\
\end{tabular}

The relatively large entropy of the gas phase products plays an
important role in determining which reaction is favored at a given
temperature.  Reactions yielding greater quantities of gaseous
products should be favored with increasing $T$, and this is consistent
with the observed trend: reaction (i), 2 mols $\rightarrow$ reaction
(c), 4 mols $\rightarrow$ reaction (f), 5 mols.  (It should be noted
that at high temperatures---and certainly above the melting point of
\foo---the harmonic approximation will no longer be valid, so we
expect some decrease in accuracy with increasing $T$.)

Of the three favorable dehydrogenation reactions, only reaction (c),
\foo\ $\rightarrow$ \libn\ + \amide\ + 4\htwo, takes place within the
temperature range where \htwo-desorption (from undoped \foo) has been
experimentally observed ($T \sim$ 520--630~K).  While our prediction
of \libn\ as a dehydrogenation product is consistent with experimental
observations, we are not aware of any reports indicating the presence
of \amide.  This discrepancy is likely due to the fact that most
experiments have been performed on off-stoichiometric \foo\ formed
from 2:1 \amide:\libh\ mixtures.  Mixtures with this composition are
deficient in the Li and N presumably needed to yield the additional
amide present in our calculations.

The calculated enthalpy for reaction (c) ranges from 11.2 to 12.8
kJ/(mol \htwo) for $T = $ 300--500~K, indicating that \htwo\ release
is a weakly endothermic process.  For the higher-temperature reaction
(f), dehydrogenation is likewise predicted to be endothermic with a
slightly larger enthalpy of $\sim$31~kJ/(mol \htwo).  Only the low
temperature reaction (i) with $\Delta H \approx $ $-$(16--19)~kJ/(mol
\htwo) is found to be exothermic.  It would be interesting to test
whether the predicted temperature-dependent decomposition sequence
[(i) $\rightarrow$ (c) $\rightarrow$ (f)] could be observed
experimentally by holding \foo\ at temperatures within each of the
reaction regimes for long times.

Based on the data presented in Fig.~\ref{fig1} and
Table~\ref{reactions}, \htwo\ desorption from \foo\ is
thermodynamically favorable at essentially all non-zero temperatures:
the negative free energies of dehydrogenation suggest that \foo\ is
metastable with respect to decomposition via one of the reactions (i),
(c), or (f).  On the other hand, experiments\cite{pinkerton05} on pure
\foo\ indicate that desorption does not take place until the
temperature is raised above 520~K. Taken together, these factors
suggest that the relatively high temperatures needed in practice for
\htwo-release are a consequence of poor kinetics, not unfavorable
thermodynamics.

An important distinction between our calculations and the experimental
dehydrogenation of undoped \foo\ concerns the phase of \foo\ present
during the dehydrogenation reaction.  As mentioned above, pure \foo\
releases hydrogen from the molten state.\cite{pinkerton05} Similarly,
\htwo\ release from other hydride phases, such as \libh, also occurs
from the molten state.\cite{frankcombe05} Due to the computational
expense associated with obtaining precise energetics for liquids using
DFT, this study, and previous studies on \libh,\cite{miwa04,
  frankcombe05} have modeled the dehydrogenation reaction as a
solid-state reaction.  Since the enthalpy of the liquid phase will be
more positive than that of the corresponding solid phase, the
calculated solid-state enthalpy of reaction (c) evaluated at the
melting point ($T_\text{mp}$), $\Delta H^{T_\text{mp}} = 12.5$ kJ/(mol
\htwo), represents an \textit{upper bound} on the enthalpy for
desorption from liquid \foo.  It will be possible to more precisely
estimate the magnitude by which $\Delta H$ is overestimated once the
latent heat of melting for \foo\ has been measured.  Until such data
become available, it nevertheless seems reasonable to conclude that
the dehydrogenation of \foo\ is either weakly endothermic, or
exothermic.  This conclusion is consistent with the failed attempts at
re-hydriding \foo\ reported thus far in the
literature,\cite{pinkerton05} as a small (positive) enthalpy of
dehydrogenation will result in a similarly small thermodynamic driving
force for re-hydriding.  On the other hand, for \foo\ doped with
Pt/Vulcan carbon,\cite{pinkerton06} dehydrogenation begins below
$T_\text{mp}$.  In this case our approach of modeling solid state
reactions more accurately captures the true phase behavior of \foo,
neglecting the possible formation of C- or Pt-containing compounds.

\section{Conclusion}
Density functional theory calculations have been employed to study the
crystal structure and the finite-temperature formation and
dehydrogenation thermodynamics of \foo.  The calculations have
resolved the discrepancies in hydrogen bond lengths between the
separate structures determined via X-ray and neutron diffraction
experiments, and suggest that the neutron data yields a slightly more
accurate description of the crystal structure. All three of the
reported experimental crystal structures relax to essentially the same
structure.

For reaction energetics, our calculations indicate that \foo\
formation is exothermic, and that the dehydrogenation enthalpy of
solid-state \foo\ is temperature-dependent: \htwo\ release is
exothermic at low temperatures, and weakly endothermic at the higher
temperatures probed by recent experiments.  A non-zero latent heat for
\foo\ melting will reduce the dehydrogenation enthalpy from the liquid
state (relative to the solid state), but the size of this effect is
still to be determined. To help clarify the identity of the phases
produced via dehydrogenation we have performed a computational search
over 17 candidate dehydrogenation reactions, and identified three
reactions having favorable thermodynamics spanning the temperature
range $T = $ 0--1000~K.  All three of these reactions exhibit a
decrease in free energy, suggesting that \foo\ is a metastable phase.
For temperatures where \htwo-desorption has been experimentally
observed, the thermodynamically-favored reaction is \foo\
$\rightarrow$ \libn\ + \amide\ + 4\htwo, with an enthalpy of
11.2--12.8 kJ/(mol \htwo).  The relatively small dehydriding
enthalpies are consistent with the failed re-hydriding attempts
reported in the literature, and suggest that hydrogen release from
\foo\ is a kinetically---rather than thermodynamically---hindered
process.

\textit{Note added in proof}---During the review of this manuscript
two new studies of \foo\ were brought to our
attention.\cite{herbst06,noritake06} In the first study Noritake and
co-workers\cite{noritake06} performed a crystal structure analysis of
\foo\ using synchrotron X-ray diffraction.  While they found external
structural parameters (bcc, space group $I2_13$, $a = 10.67$~\AA)
consistent with the two reports cited
previously,\cite{filinchuk06,chater06} the reported N--H bond lengths
were more consistent with the neutron data of
Ref.~\onlinecite{chater06} (and with our calculated N--H distances)
than the X-ray data of Ref.~\onlinecite{filinchuk06}.  In the other
recent study, Herbst and Hector\cite{herbst06} examined the electronic
structure and energetics of of \foo\ using DFT.  Our results appear to
be consistent with their findings.

\acknowledgments
  The authors thank D. Halliday, A. Sudik, and J. Yang for reviewing a
  preliminary version of this manuscript. V. Ozoli\c{n}\v{s} thanks
  the U.S. Department of Energy for financial support under grant
  DE-FG02-05ER46253.


\end{document}